\documentclass{emulateapj}
\usepackage{amsfonts,amsmath,amssymb}
\usepackage{graphicx}
\usepackage{color}
\usepackage{url}

\usepackage[backref,breaklinks,colorlinks,citecolor=blue]{hyperref} 
\usepackage[all]{hypcap} 

\newcommand{\kms}{\, {\rm km}\, {\rm s}^{-1}}

\providecommand{\farcs}{.\!\!^{\prime\prime}}
\providecommand{\fras}{.\!\!^{s}}

\begin{document}

\title{
Two Local Volume Dwarf Galaxies Discovered in 21 cm Emission: Pisces A and B

}
\shorttitle{Local Dwarfs from HI}


\author{Erik J. Tollerud\altaffilmark{1,5}, 
        Marla C. Geha\altaffilmark{1} 
        Jana Grcevich\altaffilmark{2,6}, 
        Mary E. Putman\altaffilmark{3}, 
        Daniel Stern\altaffilmark{4}
       }
       
\altaffiltext{1}{Astronomy Department, Yale University, P.O. Box 208101, New Haven, CT 06510, USA; erik.tollerud@yale.edu, marla.geha@yale.edu}
\altaffiltext{2}{Department of Astrophysics, American Museum of Natural History, Central Park West at 79th St., New York, NY 10024,
USA; jgrcevich@amnh.org}
\altaffiltext{3}{Department of Astronomy, Columbia University, New York, NY 10027, USA; mputman@astro.columbia.edu}
\altaffiltext{4}{Jet Propulsion Laboratory, California Institute of Technology, 4800 Oak Grove Drive, Mail Stop 169-221, Pasadena, CA 91109, USA; daniel.k.stern@jpl.nasa.gov}
\altaffiltext{5}{Hubble Fellow}
\altaffiltext{6}{Visiting Astronomer, Kitt Peak National Observatory, National Optical Astronomy Observatory, which is operated by the Association of Universities for Research in Astronomy (AURA) under cooperative agreement with the National Science Foundation.}

\begin{abstract}
We report the discovery of two dwarf galaxies, Pisces A and B, from a blind 21 cm \ion{H}{1} search.
These were the only two galaxies found via optical imaging and spectroscopy of 22 \ion{H}{1} clouds identified in the GALFA-HI survey as dwarf galaxy candidates.
They have properties consistent with being in the Local Volume ($<10$ Mpc), and one has resolved stellar populations such that it may be on the outer edge of the Local Group ($\sim 1 \, {\rm Mpc}$ from M31). 
While the distance uncertainty makes interpretation ambiguous, these may be among the faintest starforming galaxies known.
Additionally, rough estimates comparing these galaxies to $\Lambda$CDM dark matter simulations suggest consistency in number density, implying that dark matter halos likely to host these galaxies are primarily \ion{H}{1}-rich.
The galaxies may thus be indicative of a large population of dwarfs at the limit of detectability that are comparable to the faint satellites of the Local Group. Because they are outside the influence of a large dark matter halo to alter their evolution, these galaxies can provide critical anchors to dwarf galaxy formation models.

\end{abstract}

\section{Introduction}
\label{sec:intro}

The properties of faint dwarf galaxies at or beyond the outer reaches of the Local
Group ($1-5$ Mpc) probe the efficiency of environmentally
driven galaxy formation processes and provide direct tests of
cosmological predictions \citep[e.g., ][]{kl99ms, moo99ms, stri08commonmass, krav10satrev, kirby10, BKBK11, pontzen12, geha13}.
However, searches for faint galaxies
suffer from strong luminosity and surface brightness biases that render galaxies with $L_V \lesssim 10^6 \, L_\odot$
difficult to detect beyond the Local Group \citep{tollerud08, walsh09, hargis14}.  Because of these biases, searching for
nearby dwarf galaxies with methodologies beyond the standard optical
star count methods are essential.

This motivates searches for dwarf galaxies using the 21 cm emission line of neutral hydrogen (\ion{H}{1}).  While such searches cannot identify passive dwarf galaxies like most Local Group satellites, which lack \ion{H}{1} \citep{grcevich09, spekkens14}, they have the potential to find gas-rich, potentially starforming dwarf galaxies.  This is exemplified by the case of the Leo P dwarf galaxy, found first in \ion{H}{1} and later confirmed via optical imaging \citep{leopi, leopii}.

Here we describe two faint dwarf galaxies identified via \ion{H}{1} emission in the first data release of the Galactic Arecibo L-band Feed Array \ion{H}{1} (GALFA-HI) survey \citep{Peek11galfadr1}.
As described below, they are likely within the Local Volume ($<10$ Mpc) but just beyond the Local Group ($\gtrsim 1$ Mpc), so we refer to them as Pisces A and B.
This paper is organized as follows: in \autoref{sec:data}, we present the data used to identify these galaxies. In \autoref{sec:distance}, we consider possible distance scenarios, while in \autoref{sec:conc} we provide context and some conclusions. Where relevant, we adopt a Hubble constant of $H_0=69.3 \; {\rm km \; s}^{-1}{\rm Mpc}^{-1}$ from WMAP9 \citep{wmap9}.

\section{Data}
\label{sec:data}

The two galaxies we report on here were identified initially as cold \ion{H}{1} clouds with possibly galaxy-like properties in DR1 of the GALFA-HI survey \citep{Peek11galfadr1}.  Confirmation of these clouds as galaxies required additional optical imaging and spectroscopy, which we describe below.

\subsection{\ion{H}{1} Detection}
\label{ssec:hi}

GALFA-HI was performed with the Arecibo Observatory 305-m telescope, using the ALFA feed array and the GALSPECT spectrometer. GALFA-HI DR1 \citep{Peek11galfadr1} includes velocities $|V_{\rm LSR}| < 650 \kms$, covers 7520 square degrees of sky from $\delta = -1^\circ$ to $+38^\circ$, has a channel spacing of $0.2 \kms$, and a spatial resolution of $4'$.  The sensitivity of DR1 varies with position, but the majority of the objects cataloged would have $M_{\rm HI} < 10^6 \, M_{\odot}$ if at $1 \, {\rm Mpc}$.  The two candidate dwarfs were first found in a GALFA-HI DR1 catalog that identified \ion{H}{1} clouds with sizes $<20 '$ and velocity ${\rm FWHMs} < 35 \kms$ \citep{saul12}.  From the \citet{saul12} sample of 1964 clouds, \citet{grcevichthesis} identified 51 candidate galaxies with fluxes and sizes similar to the known gas-rich Local Group dwarf galaxies (particularly Leo T).  The two candidates presented here were also identified by \citet{saul12} as being likely galaxies because they cannot be easily associated with known high velocity cloud (HVC) complexes or Galactic gas in position-velocity space.

\subsection{Optical Imaging}
\label{ssec:opti}

We performed follow-up optical imaging of 22 of the \ion{H}{1} clouds from \citet{grcevichthesis}.  These observations were performed with the pODI instrument on the WIYN Telescope in the $g$ and $r$-band filters, with integration times of 600-1200 sec per filter per target.  Standard imaging reductions were performed by the ODI Portal, Pipeline, and Archive facility.  These include bias subtraction, flat-fielding, and alignment of individual Orthogonal Transfer Array (OTA) cells into chips. The SWarp program \citep{bertinswarp}  was used to combine the individual exposures, and DAOPHOT \citep{daophot} was used to perform PSF-fitting photometry on stars in the field.

Most of the \ion{H}{1} clouds did \emph{not} have optical counterparts with morphologies like nearby galaxies within the $\sim 4'$ GALFA-HI beam.  Those in the Sloan Digital Sky Survey \citep[SDSS, ][]{dr10} footprint show neither diffuse features like the galaxies described below, nor point source overdensities to the limit of the DR 10 catalog.  Similarly, our deeper pODI imaging showed neither overdensities nor Red Giant Branch (RGB) features in the color-magnitude diagrams (CMD) down to $r \lesssim 24$ (an RGB tip distance $ > 3 \, {\rm Mpc}$) for any of the targets we observed other than the two described below.

\begin{figure}[h!]
\begin{center}
\includegraphics[width=1\columnwidth]{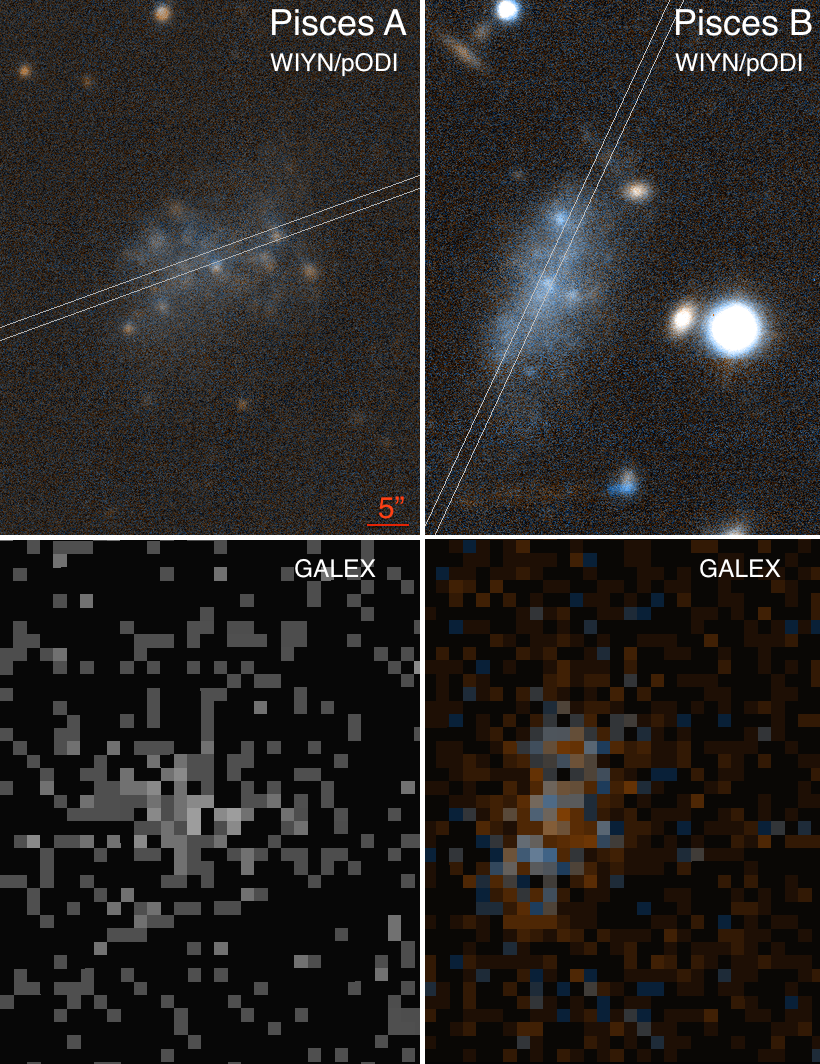}
\caption{ \protect
\label{fig:images}
Upper panels: $gr$ color composite images of dwarfs Pisces A (left) and Pisces B (right) from pODI on WIYN. The images are $1'$ tall, N is up, and E is left. The slit for the Palomar optical spectroscopy is shown as the gray lines.  Lower panels: GALEX AIS images at the same location and orientation as the upper panels.  For Pisces A (left), only NUV imaging is available, while for Pisces B (right), the image is an NUV/FUV color composite.
}
\end{center}
\end{figure}

Only two objects showed nearby dwarf galaxy-like optical counterparts within the GALFA-HI beam. The pODI images of these two candidates are shown in the upper panels of \autoref{fig:images}.  They are also visible in the SDSS, although the SDSS catalog incompletely deblends them into a mix of stars and galaxies.  Also shown in the lower panels of  \autoref{fig:images} are images from the GALEX All-sky Imaging Survey \citep[AIS, ][]{galex07}. 

The morphology of the objects in these images and presence of detectable UV flux is consistent with both being dwarf (irregular) galaxies.  Additionally, the presence of such point sources resolved in ground-based imaging implies that the galaxies are relatively nearby ($\lesssim 10$ Mpc). In particular, Pisces A (left panel of \autoref{fig:images}) shows point sources resolved enough to generate a CMD.  We discuss this further in \autoref{sec:distance} in the context of providing a distance estimate. 

While the centroid of the optical (and GALEX) objects are offset by $30-40"$ from the \ion{H}{1} emission, this is well within the $4'$ uncertainty from the GALFA-HI beam.  All other optical counterparts within the beam are less likely to be associated with the \ion{H}{1}; they either appear stellar or are consistent with being distant background galaxies (and hence at too high a redshift to match the \ion{H}{1}).  Furthermore, the H$\alpha$ emission discussed in the next section is clearly associated with these optical conterparts, and its velocity is consistent with the \ion{H}{1}, confirming the association between the optical objects and the \ion{H}{1} cloud.

\subsection{Optical Spectroscopy}

\begin{figure}[h!]
\begin{center}
\includegraphics[width=1\columnwidth]{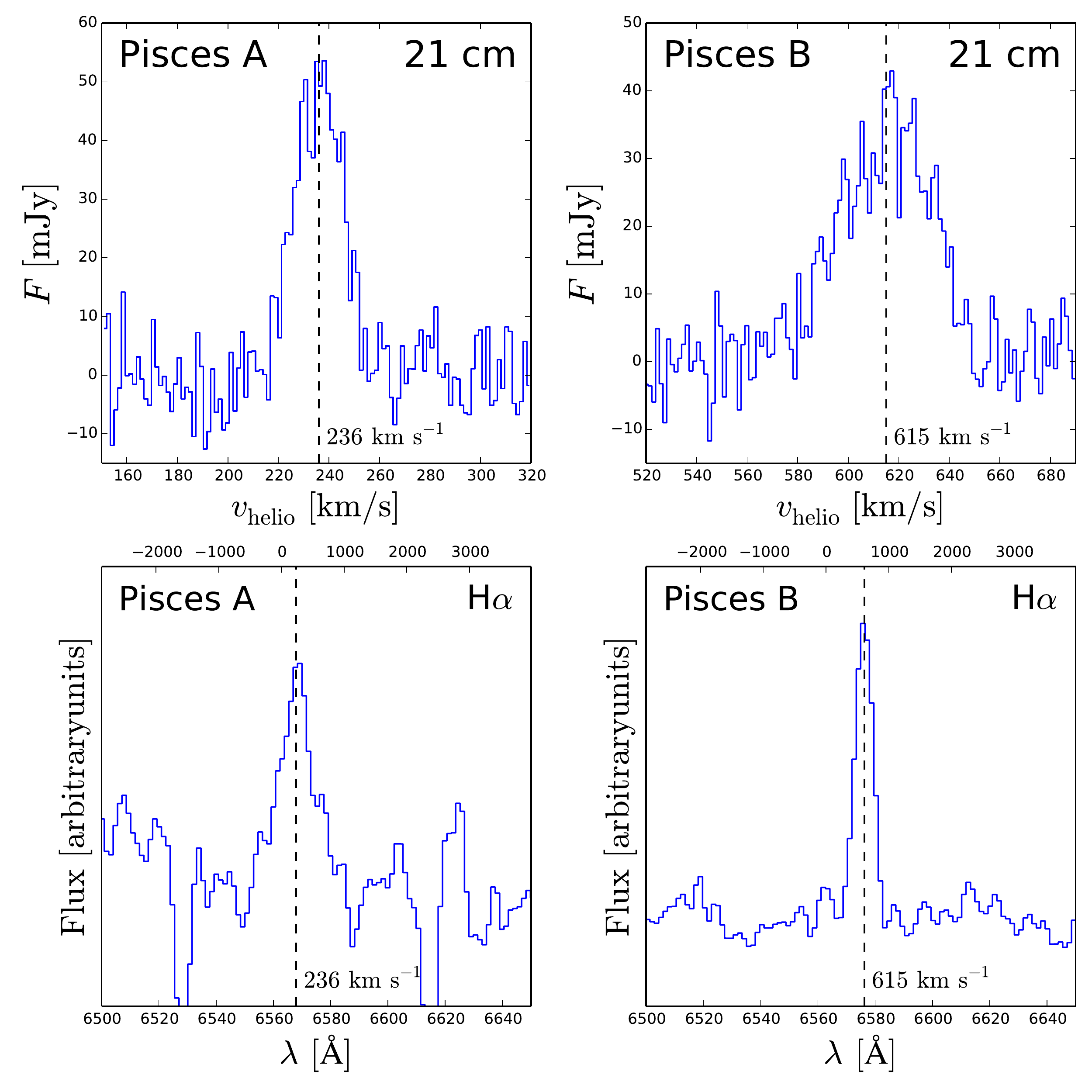}
\caption{ \protect
\label{fig:spectra}
Spectra of the dwarf galaxies. The upper panels show the GALFA-HI spectra for the clouds near Pisces A (upper-left) and Pisces B (upper-right).  The lower panels show the sky-subtracted longslit optical spectra near H$\alpha$ for Pisces A (lower-left) and Pisces B (lower-right). In all panels, the dashed black vertical lines are the emission features redshfited to $236$ and $615 \; {\rm km \; s^{-1}}$ for Pisces A and Pisces B, respectively.   Both optical spectra show H$\alpha$ emission at velocities consistent with the \ion{H}{1} peak, confirming that the optical galaxies correspond to the \ion{H}{1} clouds. The negative flux spikes at $6530$ and $6610 {\rm \AA}$ in the Pisces A optical spectrum are caused by difficult-to-subtract OH sky lines. 
}
\end{center}
\end{figure}

To confirm that the galaxies visible in \autoref{fig:images} correspond to the \ion{H}{1} clouds discussed in \autoref{ssec:hi}, we observed both galaxies in twilight with the Double Spectrograph on the Palomar Hale 200 inch telescope on UT 2014 February 6.  The slit positions are shown in the upper panels of \autoref{fig:images}.  The observations were wavelength calibrated, bias- and sky-subtracted using standard longslit techniques.  We show the spectra of these objects in \autoref{fig:spectra}, with the lower panels showing the wavelengths near H$\alpha$ and the upper panels displaying the \ion{H}{1} emission. 

The optical spectra reveal H$\alpha$ emission.  Other emission lines were not detected, but flux estimates of other lines assuming typical starforming dwarf galaxies suggest they should have ${\rm S/N} \lesssim 1$, due to poor observing conditions.  We fit the H$\alpha$ emission with Gaussian profiles, yielding a central velocity offset from the \ion{H}{1} by only $3 \pm 34$ and $10 \pm 35 \kms$ (see \autoref{tab:props}).  This is well within the H$\alpha$ $1\sigma$ uncertainties\footnote{Uncertainties on the optical velocities are much higher than for the \ion{H}{1} due to the lower resolution of the optical spectrum}, implying that the optical galaxies are indeed associated with the \ion{H}{1} clouds. Because flux calibration was not possible for these observations due to non-photometric conditions, we cannot quantify the magnitude of star formation implied by the emission.  However, the presence of any detectable H$\alpha$ emission implies star formation is ongoing (or only ceased within the last $\sim 10$ Myr) in at least some part of these galaxies.

\newcommand{\tableinforow}[1]{\multicolumn{5}{l}{\hspace{-8pt} #1 } \\}

\begin{table*}[htbp]

\begin{center}
\caption{Key observed properties of the dwarf galaxies.}
\label{tab:props}

\begin{tabular}{l  l  c  c  c c }

  \hline 
  \hline  
 & & \multicolumn{2}{ c }{Pisces A} &  \multicolumn{2}{ c }{Pisces B}    \\
 \hline 
 \hline 
 \multicolumn{6}{ c }{Distance-independent Properties} \\
 \hline 
  (1) & R.A. (J2000)                                     & \multicolumn{2}{ c }{$00^{\rm h}14^{\rm m}46\fras0 $} & \multicolumn{2}{ c }{$01^{\rm h}19^{\rm m}11\fras7 $} \\
  (2) & Dec (J2000)                                      & \multicolumn{2}{ c }{$+10^{\circ}48^{\prime}47\farcs01$} & \multicolumn{2}{ c }{$+11^{\circ}07^{\prime}18\farcs22$} \\
  (3) &  l ($^\circ$)                                    & \multicolumn{2}{ c }{108.52} & \multicolumn{2}{ c }{133.83} \\
  (4) &  b ($^\circ$)                                    & \multicolumn{2}{ c }{-51.03} & \multicolumn{2}{ c }{-51.16} \\
  (5) & $m_r$ (mag)                                      & \multicolumn{2}{ c }{$17.35 \pm 0.05$} & \multicolumn{2}{ c }{$17.18 \pm 0.07$} \\
  (6) & $(g-r)_0$ (mag)                                  & \multicolumn{2}{ c }{$0.17 \pm 0.08$} & \multicolumn{2}{ c }{$0.27 \pm 0.1$} \\
  (7) & $R_{{\rm eff}, r}$ ($^\prime$)                   & \multicolumn{2}{ c }{$7.3 \pm 0.2$} & \multicolumn{2}{ c }{$5.6 \pm  0.2$} \\
  (8) & P.A. ($^\circ$)                                  & \multicolumn{2}{ c }{$111 \pm 3$} & \multicolumn{2}{ c }{$156 \pm 1$} \\
  (9) & $F_{\rm HI}$ (${\rm Jy \; km \; s}^{-1}$)        & \multicolumn{2}{ c }{$1.22 \pm 0.07$} & \multicolumn{2}{ c }{$1.6 \pm 0.2$} \\
  (10) & $M_{\rm HI}$ ($10^5 M_{\odot} / D_{\rm Mpc}^2$) & \multicolumn{2}{ c }{$2.8 \pm 0.2$} & \multicolumn{2}{ c }{$3.8 \pm 0.4$} \\
  (11) & $v_{\rm helio, HI}$ (${\rm km \; s}^{-1}$)      & \multicolumn{2}{ c }{$236 \pm 0.5$} & \multicolumn{2}{ c }{$615 \pm 1$} \\
  (12) & $W50_{\rm HI}$ (${\rm km \; s}^{-1}$)           & \multicolumn{2}{ c }{$22.5 \pm 1.3$} & \multicolumn{2}{ c }{$43 \pm 3$} \\
  (13) & $v_{\rm helio, opt}$ (${\rm km \; s}^{-1}$)     & \multicolumn{2}{ c }{$240 \pm 34$} & \multicolumn{2}{ c }{$607 \pm 35$} \\
  (14) & $f_{\rm gas}$                                   & \multicolumn{2}{ c }{$0.80 \pm 0.02$} & \multicolumn{2}{ c }{$0.84 \pm 0.02$} \\
  \hline  
 \hline 
  \multicolumn{6}{ c }{Distance-dependent Properties} \\
 \hline 
  & Distance Scenario & Hubble flow & Nearby & Hubble flow & Nearby \\
  (15) & $D$ (Mpc) & 3.5 & 1.7 & 8.9 & 3.5 \\
  (16) & $M_r$  (mag) & -10.6 & -9.0 & -12.7 & -10.7 \\
  (17) & $R_{{\rm eff}, r}$ (pc) & 123 & 60 & 241 & 95 \\
  (18) & $M_*$ ($10^5 M_{\odot}$) & 12.2 & 2.87 & 80.8 & 12.5 \\
  (19) & $M_{\rm HI}$ ($10^5 M_{\odot}$) & 34.3 & 8.09 & 301 & 46.6 \\
  (20) & $M_{\rm tot}$ ($10^5 M_{\odot}$) & 60.2 & 14.2 & 502 & 77.7 \\
  \hline

\end{tabular}
\end{center}

Rows (1) through (14) are properties that do not require an assumed distance. (1) through (4) provide the locations of the object in J2000 Equatorial and Galactic coordinates.  Rows (5), (6), (7), and (8) are the $r$-band apparent magnitude, $g-r$ color (extinction corrected using the \citealt{sf11} correction to \citealt{sfd98}), half-light radius (on-sky), and position angle (East of North). (9) is the total \ion{H}{1} line flux, and (10) is the corresponding \ion{H}{1} mass for a galaxy at 1 Mpc.  Rows (11) and (12) are the central velocity and FWHM of the \ion{H}{1} line from GALFA-HI, and (13) gives the central velocity of the H$\alpha$ emission.  (14) is the gas fraction $f_{\rm gas} = 1.4 M_{\rm HI} / M_{\rm tot}$ (see E.g. \citealt{McGaugh12} for a discussion of the 1.4 factor). The remaining rows are properties that require an assumed distance (see \autoref{sec:distance} for a discussion of the scenarios), and the assumed distance is specified in Row (15). Rows (16) and (17) give $r$-band absolute magnitude and physical half-light radius.  Row (18) gives the stellar mass, assuming $\frac{M_*/L_V}{M_\odot/L_\odot} \sim 1$ and \citep{jester05} $r$- to $V$-band conversion. Row (19) is the \ion{H}{1} mass, and (20) is the total mass ($M_* + 1.4 M_{\rm HI}$).

\end{table*}

\section{Distances}
\label{sec:distance}

The basic details of the two objects described in \autoref{sec:data} are summarized in \autoref{tab:props}.  From these details and the morphology visible in the upper panels of \autoref{fig:images}, it is clear that they are relatively nearby dwarf galaxies.
In the imaging they are similar to the Leo P dwarf galaxy, also discovered in \ion{H}{1}.
Pisces A in particular has a very similar \ion{H}{1} line width as Leo P ($W_{50} \sim 25 \; {\rm km \; s^{-1}}$), although both it and Pisces B are optically somewhat fainter than Leo P, while having similar 21 cm flux.  
Further comparisons require the answer to a crucial question: what is the distance to Pisces A and Pisces B?  We consider this question for each galaxy in turn below.

\subsection{Pisces A}

We consider two scenarios for the distance to Pisces A.  The first is based on the assumption that Pisces A is in the Hubble flow. That is, its distance is simply $D = v/H_0 = 3.5 \, {\rm Mpc}$.  This case is considered in the first column of \autoref{tab:props}.  In this scenario, Pisces A is not part of the Local Group, but nearby, well within the Local Volume. 

Our second scenario is based on the presence of resolved stars in Pisces A (upper-left panel of \autoref{fig:images}).  The imaging shows the presence of blue point sources suggestive of young main sequence stars.  If we consider Leo P as a template for this galaxy, the brightest main sequence stars in Leo P are comparable in magnitude to the Pisces A stars (\citealt{leopii}, using the color transformations of \citealt{jester05}). Hence, for our second distance scenario, detailed in the second column of \autoref{tab:props}, we assume a distance equal to that of Leo P \citep[$1.72 \, {\rm Mpc}$, ][]{leopiv}.  Pisces A is relatively close in projection to M31, so in this scenario, $d_{\rm M31} = 1.1 \; {\rm Mpc}$.  This is comparable to the Local Group's zero-velocity distance  ($\sim 1.06 \; {\rm Mpc}$ from \citealt{mcconlgcat}), placing Pisces A just beyond the edge of the Local Group.

\subsection{Pisces B}
Pisces B has a higher radial velocity than Pisces A ($615 \; {\rm km} \; {\rm s}^{-1}$ from the \ion{H}{1}), yielding a larger distance in the Hubble flow scenario ($8.9 {\rm Mpc}$). In this case, Pisces B is still within the Local Volume, but somewhat more luminous than Leo P or Pisces A, possibly akin to a Blue Compact Dwarf (BCD).  This is further suggested by its higher \ion{H}{1} line width ($W_{50} \sim 45 \; {\rm km \; s^{-1}}$).  This scenario is considered in the third column of \autoref{tab:props}.

In the pODI imaging (upper-right panel of \ref{fig:images}), Pisces B appears to contain several potential
point sources.  However, Pisces B has more diffuse light than Pisces A,
making it difficult to obtain accurate photometry and estimate a distance based on stellar CMDs.
Additionally, the much stronger H$\alpha$ emission apparent in \autoref{fig:spectra} (lower-right panel), as well as its detection in the UV with {\it GALEX} (${\rm FUV}=18.93$, ${\rm NUV}=18.87$, Donovan Meyer et al., submitted), means that at least some of the brighter point sources may be unresolved \ion{H}{2} regions rather than distinct stars.  That said, if these point sources are resolved stars, it is possible that Pisces B is somewhat closer than the Hubble flow distance implies, and instead has a substantially positive peculiar velocity.  In the absence of deeper and higher-resolution imaging to resolve this question, in the fourth column of \autoref{tab:props}, we simply consider the \emph{limiting} case that Pisces B is as close as the Hubble Flow scenario for Pisces A.

\begin{figure}[h!]
\begin{center}
\includegraphics[width=1\columnwidth]{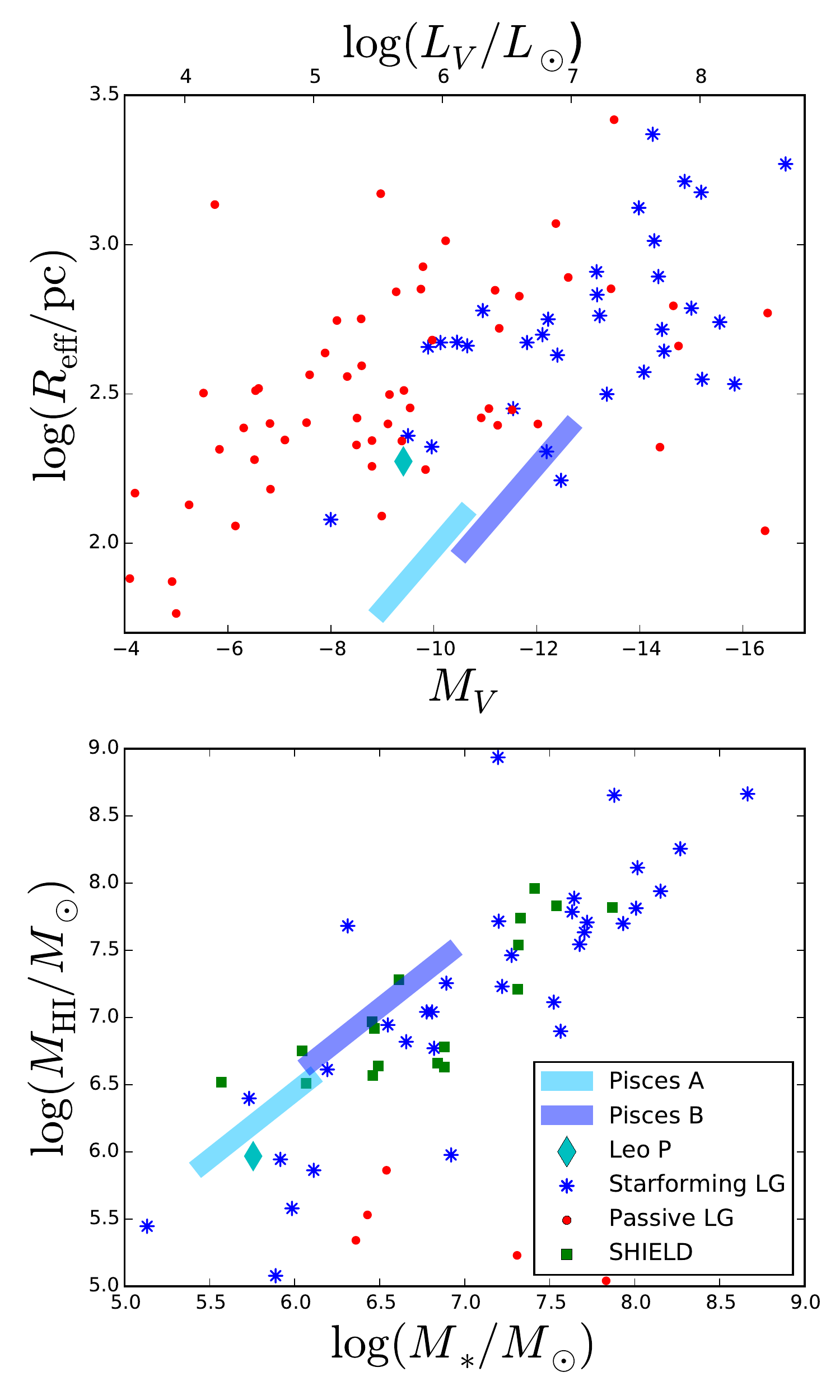}
\caption{ \protect
\label{fig:context}
Upper panel: Comparison of Pisces A and Pisces B to nearby dwarf galaxies in size vs. luminosity.  Lower panel: Comparison of Pisces A and Pisces B to nearby galaxies in stellar mass vs. \ion{H}{1} mass.  Local Group galaxy properties are from \citet{mcconlgcat}. Leo P properties are from \citet{leopiv} (the size described there is for an outer detectable extent, and thus is an upper limit on $R_{\rm eff}$).  The SHIELD sample of low-mass \ion{H}{1}-rich nearby dwarfs is from \citet{SHIELD}.  The shaded bars show the properties of Pisces A (cyan) and Pisces B (blue) between the two distance scenarios discussed in the text.  For the upper panel, we use the \citet{jester05} transformations to convert from $r$-band magnitudes for Pisces A and Pisces B.  For the lower panel, we make the estimate $\frac{M_*/L_V}{M_\odot/L_\odot} \sim 1$.
}
\end{center}
\end{figure}

\section{Discussion and Conclusions}
\label{sec:conc}

In \autoref{fig:context}, we show Pisces A and B in the context of nearby dwarf galaxies. These include the Local Group dwarfs (as compiled in \citealt{mcconlgcat}), the SHIELD sample of nearby $10^6 - 10^7 \, M_{\odot}$ \ion{H}{1}-rich galaxies  \citep{SHIELD}, and Leo P.  The upper panel demonstrates that, depending on the distance, the galaxies described here may be among the faintest known star-forming galaxies.  They also overlap in their basic structural properties with Local Group dwarfs, although they are somewhat more compact.  Unlike the Local Group galaxies, however, they are well beyond the virial radii of any large dark matter halo like that of the MW or M31.  Hence, they are crucial data points both for understanding how star formation functions at the lowest luminosities and as possible progenitors of the faint (predominantly passive) dwarf satellites of the Local Group.

Additionally, the lower panel of \autoref{fig:context} shows they have a slightly higher \ion{H}{1} mass relative to typical dwarfs of the same luminosity.  The resulting gas fractions (last row of \autoref{tab:props}), while high, are not necessarily surprising, as both the mean and scatter in gas fractions increase for fainter dwarfs (\citealt{geha06}, Bradford et al. in prep).  Furthermore, these galaxies were discovered initially via \ion{H}{1}, so detection biases favor higher \ion{H}{1} masses. Nevertheless, the existence of a significant nearby population of faint galaxies with high \ion{H}{1} content may offer significant constraints on dwarf galaxy formation models.

Motivated by the existence of these galaxies, we are identifying further candidate nearby, faint, starforming dwarfs in the SDSS.  As discussed in \autoref{ssec:opti}, Pisces A and Pisces B have blue point sources in the SDSS catalog (although it mis-classifies them as galaxies due to the surrounding diffuse light).  We conducted a search in the DR10 catalog for similar clusters of blue objects not associated with known galaxies or Galactic structures.  While most such overdensities turn out to be artifacts, $\sim 100$ show morphologies potentially consistent with nearby dwarf galaxies.  An ongoing spectroscopic follow-up campaign on these objects shows that some do indeed have H$\alpha$ emission consistent with local galaxies, and we will present these results in a future paper.

\subsection{Comparison to Simulation}

Detailed use of these galaxies as data points for galaxy formation will require more firm distances, only possible with deeper and/or higher resolution photometry (which will be obtained by our approved Cycle 22 {\it Hubble Space Telescope} program).  However, the detection of Pisces A may have interesting interpretations in a cosmological context.  To demonstrate this, we consider simulations from the ELVIS suite, designed to resemble the Milky Way/M31 pairing of the real Local Group \citep{gk14elvis}.  We mock observe dark matter halos in a frame with $v_{\rm tan}=220 \kms$ and $d=8 \; {\rm kpc}$ (relative to the host halo). GALFA-HI can distinguish high velocity cloud complexes from galaxy candidates at $v_{\rm helio} \gtrsim 90 \kms$ \citep{saul12}, so we only consider halos with corresponding mock $v_{\rm helio}$.  We also ignore halos \emph{inside} the host's virial radius, and with stellar masses\footnote{Stellar masses in ELVIS were assigned via the abundance matching prescription of \citet{gk14elvis}.} $10^5 <M_*/M_\odot < 10^8$.  This yields $13 \pm 4$ halos.

The first data release of GALFA-HI  covers $18\%$ of the sky, so we expect of order $1-3$ galaxies to be found in GALFA-HI based on the above estimate of the number of ELVIS halos.  If Pisces A is anywhere in the distance range we consider here, it is the only one in the GALFA-HI DR1 that falls in that range. (The most likely distances for Pisces B place it beyond the edge of most of the ELVIS simulations.)  This implies there is at least not an order-of-magnitude discrepancy between the expectations of $\Lambda$CDM and the observations.  While only one object, this also hints that because the number density of \ion{H}{1}-selected faint dwarfs is roughly the same as that of corresponding ELVIS halos, even these low-mass halos typically have \ion{H}{1} if they are beyond the Local Group.

\subsection{Conclusions}
 We report two new nearby dwarf galaxies discovered from their \ion{H}{1} emission, Pisces A and B.  They provide new opportunities to study some of the faintest starforming galaxies.  While detailed interpretation of the galaxies depends on the (uncertain) distance, rough estimates of number density suggest broad consistency with $\Lambda$CDM.  They also suggest that these galaxies could represent a larger population of nearby faint starforming dwarfs at the limits of detectability. These galaxies, and others like them, may thus provide a stepping stone from the dwarfs of the Local Group to the realm beyond.

\vspace*{1.5 \baselineskip}

We gratefully acknowledge Roberto Assef and Alessandro Rettura for assisting with the Palomar observations.  We also acknowledge Lauren Chambers and Adrian Gutierrez for obtaining additional 21 cm observations of Pisces A that improved interpretation of the GALFA-HI spectra. We further acknowledge Josh Peek and Jen Donovan Meyer for helpful discussions.  Additionally, we acknowledge the anonymous referee for helpful suggestions that improved this paper.

This research made use of Astropy, a community-developed core Python package for Astronomy \citep{astropy}.  It also used the MCMC fitting code {\it emcee} \citep{emcee}.

Support for EJT was provided by NASA through Hubble Fellowship grant \#51316.01 awarded by the Space Telescope Science Institute, which is operated by the Association of Universities for Research in Astronomy, Inc., for NASA, under contract NAS 5-26555. MEP  acknowledges support by NSF grant \#AST-1410800.

This work is based on observations obtained at the WIYN Telescope. The WIYN Observatory is a joint facility of the University of Wisconsin-Madison, Indiana University, Yale University, and the National Optical Astronomy Observatory

This work is based on observations obtained at the Hale Telescope,
Palomar Observatory as part of a continuing collaboration between
the California Institute of Technology, NASA/JPL, Oxford University,
Yale University, and the National Astronomical Observatories of
China.

\end{document}